\documentclass[graybox, envcountchap]{SNmult}

\usepackage{type1cm}		

\usepackage{makeidx}		 
\usepackage{graphicx}		
\usepackage{multicol}		
\usepackage[bottom]{footmisc}
\usepackage{newtxtext}	   
\usepackage[varvw,varbb]{newtxmath}	   

\usepackage[utf8]{inputenc}
\usepackage[T1]{fontenc} 
\usepackage[english]{babel}
\usepackage{fix-cm} 
\usepackage{microtype} 
\usepackage{chapterbib} 
\usepackage{etoolbox} 
\AtBeginDocument{\patchcmd{\bibsection}{\addcontentsline{toc}{section}{\refname}}{}{}{}}
\usepackage{amsthm,amsfonts} 
\usepackage{isotope}
\usepackage{braket}
\usepackage{siunitx}
\DeclareSIUnit\torr{Torr} 
\DeclareSIUnit\gauss{G}
\DeclareSIUnit\atoms{atoms}

\makeatletter
\providecommand*{\toclevel@titlech}{0} 
\edef\toclevel@authorch{\the\numexpr\toclevel@titlech+1}
\providecommand*{\toclevel@titlech}{0}
\def\toclevel@authorch{1000}
\makeatother

\usepackage{hyperref}
\usepackage[hyperpageref]{backref} 
\hypersetup{
    colorlinks=true,
    linkcolor=blue,
    filecolor=blue,      
    urlcolor=cyan,
    citecolor=blue,
    pdftitle={Synthetic Polariton Matter in the solid state},
    pdfauthor={Sylvain Ravets},
    pdfpagemode=FullScreen,
    }

\makeindex             

\begin{document}

\title*{Synthetic Polariton Matter in the solid state}
\author{Sylvain Ravets\orcidID{0000-0003-1746-5245}}
\institute{Sylvain Ravets \at Center for Nanoscience and Nanotechnology, CNRS, Paris–Saclay University, 91120 Palaiseau, France, \email{sylvain.ravets@c2n.upsaclay.fr}}

\maketitle

\abstract*{Synthetic materials are obtained by assembling atoms or artificial atoms into regular arrays, thereby forming artificial crystals that offer powerful platforms to emulate and explore condensed-matter phenomena in highly controlled settings. They enable probing outstanding questions in many-body physics and designing new phases of matter with no direct analogue in nature. Beyond their fundamental interest, these materials hold potential for future technological applications through the emergence of novel concepts and functionalities. Synthetic materials have been engineered using a wide range of physical platforms, including both natural atoms and fabricated artificial atoms in the solid-state. A particularly intriguing approach relies on photons. When confined in optical cavities and strongly coupled to matter excitations, photons acquire an effective mass and can experience interactions, giving rise to hybrid light-matter quasiparticles known as polaritons. By arranging polaritons in periodic structures, one can engineer synthetic photonic materials with tailored band structures and controllable interactions, offering a promising route toward exploring strongly correlated photonic phases. This chapter focuses on a solid-state realization of such systems: exciton polaritons confined in semiconductor microcavities. Following a general introduction, we describe how photon mass and photonic band structures emerge from cavity confinement, and how interactions arise via strong coupling to excitons in quantum wells. We finally review how these ingredients can be used to explore rich physics from the mean-field to the quantum regime.}

\abstract{Synthetic materials are obtained by assembling atoms or artificial atoms into regular arrays, thereby forming artificial crystals that offer powerful platforms to emulate and explore condensed-matter phenomena in highly controlled settings. They enable probing outstanding questions in many-body physics and designing new phases of matter with no direct analogue in nature. Beyond their fundamental interest, these materials hold potential for future technological applications through the emergence of novel concepts and functionalities. Synthetic materials have been engineered using a wide range of physical platforms, including both natural atoms and fabricated artificial atoms in the solid-state. A particularly intriguing approach relies on photons. When confined in optical cavities and strongly coupled to matter excitations, photons acquire an effective mass and can experience interactions, giving rise to hybrid light-matter quasiparticles known as polaritons. By arranging polaritons in periodic structures, one can engineer synthetic photonic materials with tailored band structures and controllable interactions, offering a promising route toward exploring strongly correlated photonic phases. This chapter focuses on a solid-state realization of such systems: exciton polaritons confined in semiconductor microcavities. Following a general introduction, we describe how photon mass and photonic band structures emerge from cavity confinement, and how interactions arise via strong coupling to excitons in quantum wells. We finally review how these ingredients can be used to explore rich physics from the mean-field to the quantum regime.}

\section{Synthetic matter: state of the art}
\label{Ravets:sec:1}

Our focus is on two-dimensional (2D) synthetic materials, constructed by assembling regular arrays of coupled particles into artificial crystals. In what follows, we first mention the general context of real 2D materials found in nature, which have been a great source of inspiration for this field. We then discuss synthetic 2D crystals and identify the key ingredients required to construct them. We finally introduce the specific case of photonic synthetic matter.

\subsection{Regular atom arrays: from atomic monolayers to synthetic materials}

One of the major achievements in modern condensed matter physics was the discovery that genuine two-dimensional (2D) crystals can be isolated from bulk materials~\cite{Novoselov2004, Novoselov2005}. Graphene, obtained by exfoliation from graphite, was the first such material, and its discovery has inspired a vast body of research~\cite{Geim2007, CastroNeto2009}. Since then, a broad family of 2D compounds has been studied, including hexagonal boron nitride (h-BN) and transition metal dichalcogenides (TMDs), often assembled into van der Waals heterostructures~\cite{Geim2013, Novoselov2016, Wang2018}.

The electronic band structures of these systems, arising from the hybridization of atomic orbitals, determine many of their fundamental properties~\cite{Ashcroft1976, Kittel2005}: whether they are metals, insulators, or semimetals. Beyond this single-particle picture, electronic interactions enrich the physics even further, leading to strongly correlated many-body phases. For example, fractional quantum Hall states have been observed in graphene under strong magnetic fields~\cite{Du2009, Bolotin2009}, and superconductivity has been found in carefully engineered van der Waals heterostructures~\cite{Cao2018}. Such states have been probed through transport measurements, the standard diagnostic tool in solid-state condensed matter, which have led to remarkable advances. Nevertheless, many fundamental questions remain open, and new approaches to design, probe and control many-body quantum phases are highly sought after.

In parallel with these advances, intense efforts have been devoted to the bottom-up assembly of synthetic 2D materials, in which atoms or artificial atoms are coupled together to form ordered structures~\cite{Feynman1982, Bloch2008, Blatt2012, Houck2012, Carusotto2013, Gross2017, Carusotto2020, Browaeys2020, Altman2021, Grass2025}. These platforms offer complementary opportunities to the study of natural crystals: their characteristic energy and length scales are widely different, and they allow direct access to observables that are challenging to measure in real materials. Of particular interest are local observables such as spatially resolved spectroscopy of crystal orbitals or defect modes, combined real- and reciprocal- space imaging, measurements of spatio-temporal correlation functions, and dynamics triggered by local quenches.

This field is motivated by both fundamental and applied considerations. On the fundamental side, synthetic materials enable the realization of simplified model Hamiltonians, directly inspired by theoretical proposals. They allow one to vary the lattice geometry at will and to introduce different ingredients one at a time, such as spin-orbit coupling, time-reversal symmetry breaking, controlled defects, external pumping, or engineered dissipation. On the applied side, these platforms provide opportunities to harness exotic states of matter for technology, with potential applications ranging from device physics (novel light sources, photonic devices), to quantum-enhanced sensing and quantum technologies.

\subsection{Synthetic photonic materials?}

\begin{figure}[ht]
    \centering
	\includegraphics[width=0.9\textwidth]{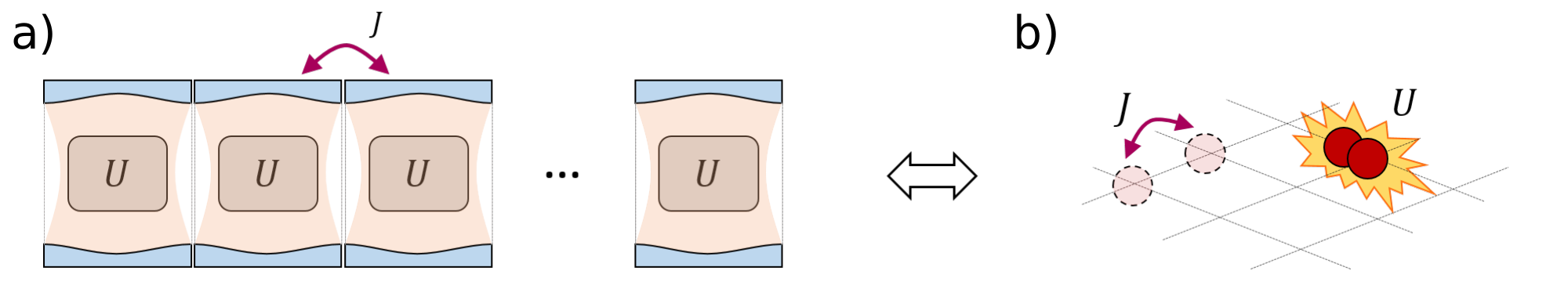}
	\caption{Schematic illustration of a synthetic photonic material composed of an array of coupled optical cavities. \textbf{(a)} Neighboring cavities are coupled with strength $J$, while each cavity contains a nonlinear medium that induces an effective photon–photon interaction $U$. \textbf{(b)} The system maps onto a Bose–Hubbard model with hopping amplitude $J$ and on-site interaction $U$.}
	\label{Ravets:fig:1}
\end{figure}

The assembling of synthetic materials generally requires a few key ingredients. First, one needs an elementary building block that plays the role of an artificial atom, providing localized orbitals. Second, these building blocks must be coupled together to form periodic arrays, thereby forming single-particle band structures. Finally, on-site interactions are required to generate quantum correlations and many-body phenomena. These principles have been implemented in a broad variety of experimental platforms, including cold atoms trapped in optical lattices~\cite{Bloch2008, Gross2017}, trapped ions in electromagnetic potentials~\cite{Blatt2012}, Rydberg atoms~\cite{Browaeys2020}, as well as artificial atoms in the solid state such as semiconductor quantum dots~\cite{Hensgens2017}, color centers~\cite{Atature2018}, and microwave or optical photons in cavities~\cite{Houck2012, Carusotto2020, Hartmann2008, Hartmann2016, Noh2017}.

Here, we concentrate on the latter approach: synthetic materials built from photons. At first sight, this idea seems counterintuitive. Photons indeed differ fundamentally from electrons: they are massless in free space, they interact only very weakly, and obey bosonic statistics. Nonetheless, approaches have been proposed to realize quantum phases of light in synthetic crystalline materials, by confining photons into arrays of nonlinear cavities~\cite{Hartmann2006, Greentree2006, Angelakis2007}, as illustrated in Fig.~\ref{Ravets:fig:1}. The essential ingredients at the core of these ideas are the following:
\begin{enumerate}
\item Cavity confinement endows photons with an effective mass, allowing them to behave as massive particles, 
\item By engineering a coupling $J$ between neighboring cavities, one can realize tailored photonic band structures,
\item embedding nonlinear media within the cavities, one can lead to a ``photon blockade effect'' and induce effective photon–photon interactions $U$~\cite{Tian1992, Imamoglu1997, Birnbaum2005, Verger2006}.
\end{enumerate}
Together, these elements provide all the necessary ingredients to realize synthetic photonic materials, enabling the realization of correlated many-body bosonic states of light such as those described by the Bose–Hubbard model~\cite{Hartmann2006, Greentree2006, Angelakis2007, Carusotto2009, Carusotto2013, LeBoite2013} or bosonic analogues of quantum Hall states~\cite{Cho2008, Hafezi2011, Umucalilar2011, Umucalilar2012, Hafezi2013, Clark2020}. In the following sections, we will explore how exciton–polaritons in semiconductor microcavities provide a particularly powerful realization of this idea, offering both the ingredients for synthetic matter and the ability to probe it with exquisite experimental control.

\section{Band structure engineering in arrays of coupled microcavities}
\label{Ravets:sec:2}

In this section, we explore how cavity photons can be used as building blocks to realize photonic materials. When confined in an optical cavity, photons acquire an effective mass and behave as massive particles. Nanofabricated microcavities confining photons in all spatial directions support discrete photonic modes that can be regarded as artificial atomic orbitals. By further arranging such cavities into ordered arrays with controlled coupling, one can engineer photonic band structures.

\subsection{Longitudinal confinement: providing mass to photons}
\label{Ravets:subsec:2.1}

\begin{figure}[htb]
    \centering
	\includegraphics[width=0.9\textwidth]{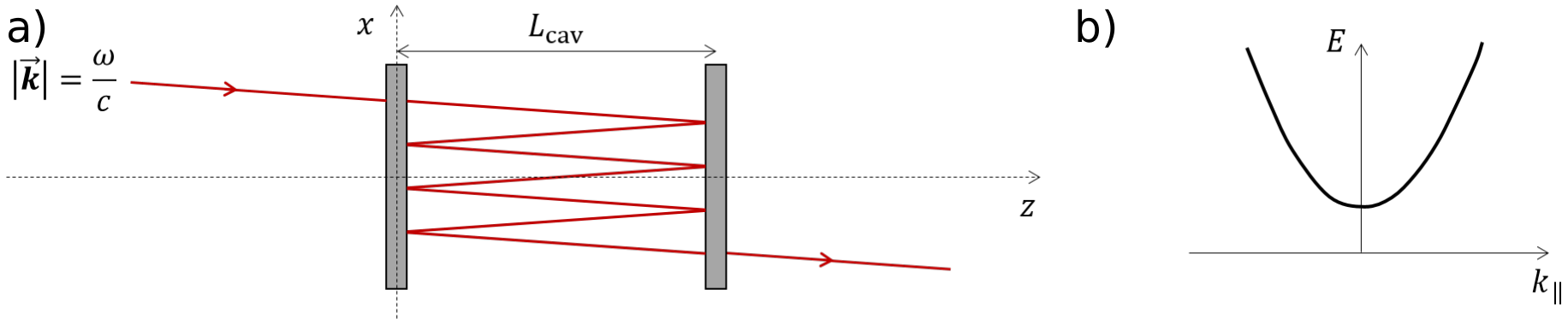}
	\caption{Geometrical optics illustration of a planar Fabry–-Perot cavity. \textbf{(a)} A light ray entering the cavity undergoes multiple reflections between the mirrors before escaping, during which it experiences a transverse drift that depends on the in-plane wavevector $\mathbf{k}_{\parallel}$. \textbf{(b)} This transverse motion is equivalent to that of a massive particle, as reflected in the cavity photon dispersion relation: the energy varies quadratically with $\left| \mathbf{k}_{\parallel} \right|$, providing the confined photons with an effective mass.}
	\label{Ravets:fig:2}
\end{figure}

We consider a Fabry--Perot cavity formed by two parallel mirrors separated by a distance $L_0$, thus defining a cavity that is filled with a material of refractive index $n$ (Fig.~\ref{Ravets:fig:2}a). A plane wave with wavevector $\mathbf{k} = \mathbf{k}_{\parallel} + k_z \mathbf{e_z}$, bouncing between the mirrors, accumulates a phase $2k_z L_0$ during each round trip. As a consequence of the photon confinement along one direction, the cavity resonance condition imposes the quantization of the longitudinal wavevector:
\begin{equation}
2k_z L_0 = 2 \pi p 
\quad \Rightarrow \quad 
k_z = \frac{p \pi}{L_0} \;,
\end{equation}
with $p$ an integer labeling the longitudinal mode.

We first consider a geometrical optics picture. Fig.~\ref{Ravets:fig:2}a shows a ray entering the cavity at an angle $\theta$. Within the cavity, the ray undergoes multiple reflections before eventually leaking out, as a consequence of the finite reflectivity of the mirrors (finite photon lifetime in the cavity). During each round trip, the photon experiences a transverse drift, whose magnitude directly depends on the incidence angle (or equivalently on the in-plane momentum $\mathbf{k}_\parallel$). This contrasts to the case of photons propagating in free space, where the velocity $c/n$ does not depend on $\mathbf{k}$. Here, the transverse velocity depends on the wavevector, which is the expected behavior for massive particles.

To quantify this effect, we evaluate the transverse velocity $v_{\parallel}$. In the paraxial approximation ($|\mathbf{k}_{\parallel}| \ll k_z$), the incidence angle can be expressed as:
\begin{equation}
\theta \simeq \frac{\left| \mathbf{k_{\parallel}}\right| }{ k_z } = \frac{L_0 {k}_{\parallel}}{p \pi} \;. 
\end{equation}
After a round trip of duration $dt = 2nL_0/c$, the photon undergoes a transverse shift $dx \simeq 2L_0 \theta$, and therefore the resulting transverse velocity reads:
\begin{equation}
v_{\parallel} = \frac{dx}{dt} = \frac{c L_0}{n \hbar p \pi} \hbar k_{\parallel} \;.
\end{equation}
This expression is exactly that of a massive particle, with effective mass:
\begin{equation}
m_{\mathrm{ph}} = \frac{n \hbar p \pi}{c L_0} \;.
\label{Ravets:eq:PhotonMass}
\end{equation}
In other words, photons vertically confined in a planar cavity behave, in their transverse dynamics, as massive particles.

The same conclusion can be reached from the photon dispersion relation in the cavity:
\begin{equation}
\hbar \, \omega(\mathbf{k}) = \frac{\hbar c}{n} |\mathbf{k}| \;.
\end{equation}
Inserting the quantization of $k_z$ and expanding in the paraxial limit,
\begin{equation}
\hbar \, \omega(\mathbf{k}) = \frac{\hbar c}{n} \sqrt{ \left( \frac{p \pi}{L_0} \right)^2 + k_{\parallel}^2 } 
\simeq \hbar \omega_0 + \frac{\hbar^2 k_{\parallel}^2}{2 m_{\mathrm{ph}}} \;,
\end{equation}
where $\hbar \omega_0 = m_{\mathrm{ph}} (c/n)^2$. We recover the parabolic dispersion typical of a particle of mass $m_{\mathrm{ph}}$ (Fig.~\ref{Ravets:fig:2}b), with its kinetic energy term equal to $m_{\rm ph} v_{\parallel}^2 /2$. For reference, the effective photon mass in a microcavity of length $L_0 \sim \qty{1}{\micro \meter}$ is typically $m_{\rm ph} \sim 10^{-5} m_{e^-}$, where $m_{e^-}$ is the electron mass. This is several orders of magnitude lighter than the mass of typical electronic excitations in materials, which is of order $m_{e^-}$.

\subsection{Lateral confinement: atomic orbitals for photons}
\label{Ravets:subsec:2.2}

\begin{figure}[htb]
    \centering
	\includegraphics[width=0.9\textwidth]{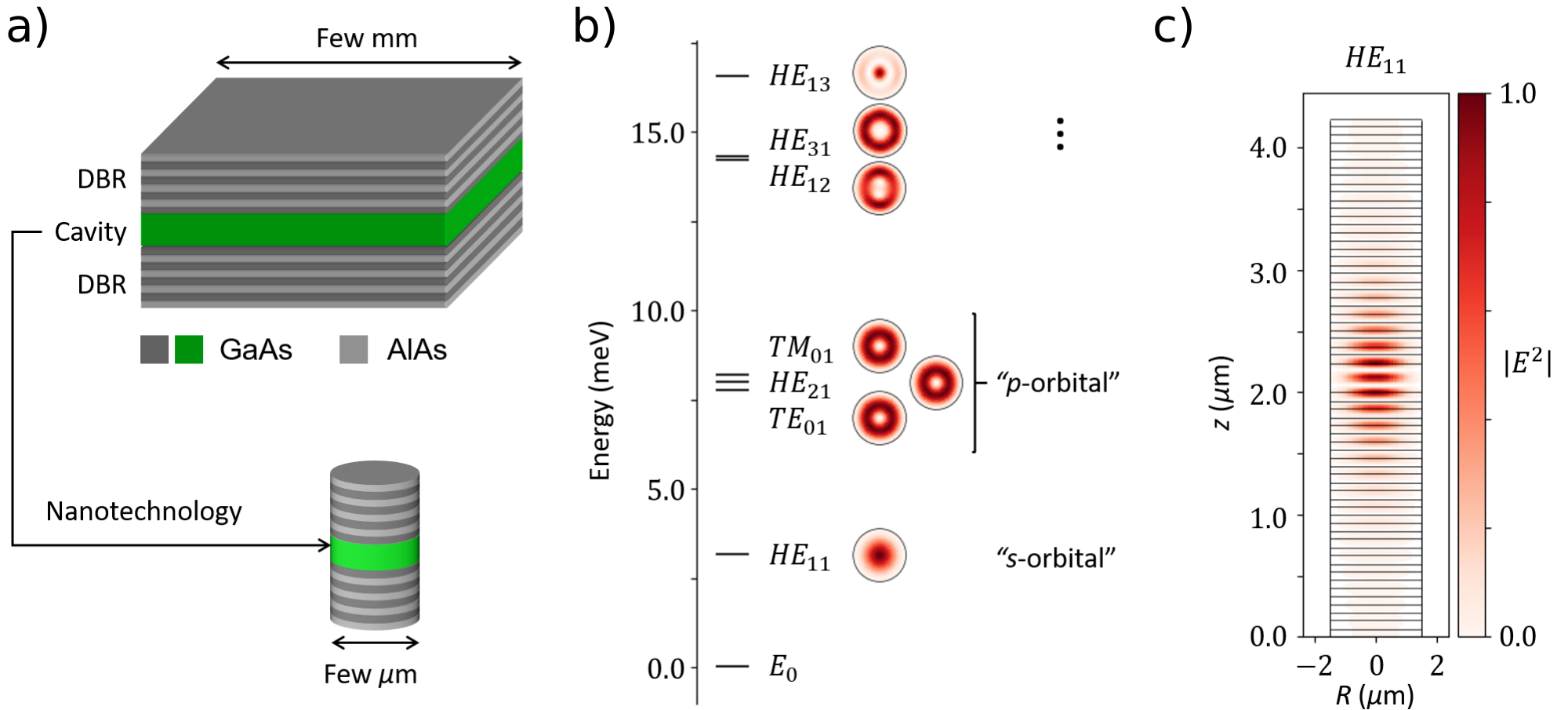}
	\caption{Photon confinement in semiconductor microcavities. 
	\textbf{(a)} Schematic of a Bragg cavity composed of two distributed Bragg reflectors (grey) and a central cavity spacer (green). Using nanofabrication techniques, lateral microstructures can be defined, typically a few microns in diameter. 
	\textbf{(b)} Calculated energy spectrum for a $\qty{3}{\micro\meter}$ micropillar, obtained by numerically solving the full three-dimensional Maxwell equations for a structure containing 15 Bragg mirror pairs in both the top and bottom DBR, and a cavity spacer of optical thickness $\lambda_0$ (design wavelength $\lambda_0 = \qty{852}{\nano\meter}$). A series of discrete modes emerges, labeled according to the hybrid electric (${\rm HE}_{nm}$) modes of a cylindrical waveguide. The energy origin $E_0$ corresponds to the planar cavity mode at $\mathbf{k}=0$. Insets show the normalized transverse intensity profiles (modulus squared of the electric field). 
	\textbf{(c)} Longitudinal cut of the normalized intensity for the fundamental ${\rm HE}_{11}$ mode, showing the standing-wave pattern within the cavity spacer and Bragg mirrors.}
	\label{Ravets:fig:3}
\end{figure}

For exciton–polaritons, the standard platform consists of monolithic planar Bragg microcavities based on semiconductor heterostructures. These structures are fabricated by molecular beam epitaxy (MBE), a technique that enables crystals to be grown with atomic-layer precision, providing exquisite control over thickness, composition, and purity. A widely used material system is the alloy family ${\rm Al}_x{\rm Ga}_{1-x}{\rm As}$. By tuning the aluminum content $x$, one can span the full range from ${\rm AlAs}$ to ${\rm GaAs}$. Remarkably, these two compounds share nearly identical lattice constants, which permits the growth of high-quality heterostructures with atomically sharp interfaces and minimal strain, even over many repeated layers.

\subsubsection{Semiconductor planar microcavities}

To fabricate optical cavities, one uses the fact that different ${\rm Al}_x{\rm Ga}_{1-x}{\rm As}$ alloys have different dielectric properties, as can be seen from their different refractive indices $n_{\rm AlAs} \simeq 2.90$ and $n_{\rm GaAs} \simeq 3.45$. The cavity mirrors are distributed Bragg reflectors (DBRs), consisting of alternating layers of ${\rm AlAs}$ and ${\rm GaAs}$. At each interface, Fresnel reflection occurs due to the refractive index contrast, so that the multiple reflected waves interfere with one another. When the layer thicknesses $d_{\rm GaAs}$ and $d_{\rm AlAs}$ are chosen to satisfy the Bragg condition:
\begin{equation}
n_{\rm GaAs} \, d_{\rm GaAs} = n_{\rm AlAs} \, d_{\rm AlAs} = \lambda_0/4 \;,
\end{equation}
constructive interference occurs over a spectral range centered around the design wavelength $\lambda_0$, where the reflectivity is maximum. As these mirrors rely on interference effects, high reflectivities can be reached: a DBR stack of about thirty DBR pairs typically exceeds $99.995~\%$. Finally, by growing a $\rm GaAs$ spacer in between two DBR mirrors, one defines a planar Fabry–Perot microcavity that confines light along the growth axis (Fig.~\ref{Ravets:fig:3}a. The cavity spacer thickness is $L_{\rm cav}$ determines the resonance wavelength of the cavity and is chosen equal to $n_{\rm GaAs} L_{\rm cav} = \lambda_0$.

As discussed in Sect.~\ref{Ravets:subsec:2.1}, the planar cavity exhibits a parabolic dispersion relation, $\omega(\mathbf{k}) = \omega_0 + \hbar \mathbf{k}^2/(2m_{\rm ph})$. In Bragg cavities, where light propagates through alternating layers with different refractive indices, it is useful to introduce an effective refractive index $n_{\rm eff}$ that represents the average optical environment experienced by photons. This quantity can be defined using the expression of the effective photon mass given in Eq.~\ref{Ravets:eq:PhotonMass}: 
\begin{equation}
n_{\rm eff} = c \, \sqrt{\frac{\hbar \omega_0}{ m_{\rm ph}}} \;.
\label{Ravets:eq:neff}
\end{equation}
This concept will be particularly useful in Sect.~\ref{Ravets:subsec:2.3} when reducing the problem to a two-dimensional description.

\subsubsection{Confining photons in micropillar cavities}

Planar cavities confine light only along the growth direction, so additional patterning is needed to achieve full three-dimensional confinement and create discrete, localized modes analogous to atomic orbitals, which will serve as the fundamental building blocks for photonic materials. Several nanofabrication techniques exist to achieve lateral confinement~\cite{Amo2016, Schneider2017}. Here, we discuss the ``deep etching'' method, in which the cavity spacer and mirrors are patterned into micropillars by etching down to the substrate, with electron-beam lithography and dry etching as the primary fabrication steps. This creates a step-index boundary between the high-index semiconductor and vacuum, leading to total internal reflection and lateral confinement, similar to a step-index optical fiber. For micropillars with diameters of only a few microns (comparable to the optical wavelength of light), the cavity thus supports a discrete set of confined optical modes. These confined modes are eigenmodes of the electromagnetic field in the full three-dimensional micropillar structure, and are obtained as solutions of Maxwell’s equations:
\begin{equation}
\nabla \times \nabla \times \mathbf{E}(\mathbf{r}) - \left( \frac{\omega}{c} \right)^2 n^2(\mathbf{r}) \, \mathbf{E}(\mathbf{r}) = 0 \;.
\label{Ravets:eq:Maxwell3D}
\end{equation}

Numerical solutions of the full three-dimensional Maxwell equations indeed reveal a discrete energy spectrum (Fig.~\ref{Ravets:fig:3}b, with each energy level corresponding to a cavity confined mode. These modes can be classified as hybrid electric (${\rm HE}_{nm}$) modes of a cylindrical waveguide~\cite{Yariv1989, Panzarini1999}. Fig.~\ref{Ravets:fig:3}b-c show the calculated electric field distribution in both the longitudinal and transverse planes for the lowest energy (fundamental) ${\rm HE}_{11}$ mode. The longitudinal profile clearly reflects the cavity structure: the field exhibits standing-wave oscillations within the structure and a field maximum at the cavity center due to mirror confinement (Fig.~\ref{Ravets:fig:3}c. The transverse profile in Fig.~\ref{Ravets:fig:3}b highlights the lateral confinement and exhibits a single-lobe distribution, which we identify as the photonic analogue of hydrogen atom $s$-orbital. In the rest of this work, we concentrate on this lowest-energy ``photonic $s$-orbital'' and discuss the formation of photonic energy bands through $s$-orbital hybridization. Note that in actual microcavity and structures, higher-order transverse modes are always present. Although they are usually well separated from the $s$ modes and can often be neglected in a first approximation, a fine quantitative understanding of mode hybridization may require explicitly accounting for these higher-order modes.

\subsection{Photonic band structures}
\label{Ravets:subsec:2.3}

\begin{figure}[ht]
    \centering
	\includegraphics[width=0.9\textwidth]{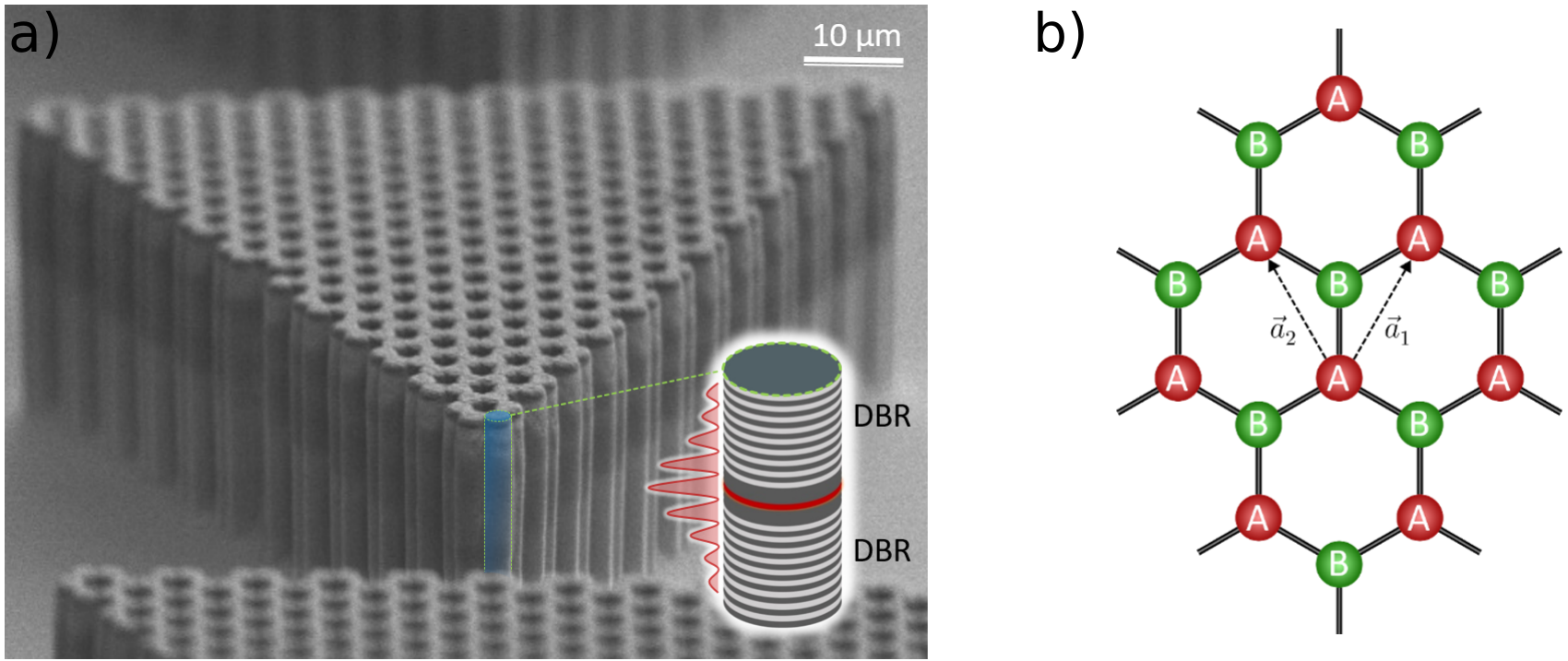}
	\caption{A regular lattice of polariton micropillars. 
	\textbf{(a)} Scanning electron microscope image of a honeycomb lattice of polariton micropillars. The inset shows a single micropillar used as a building block. 
	\textbf{(b)} Schematic of the honeycomb lattice, highlighting the two sublattices $A$ and $B$ and the primitive lattice vectors $\mathbf{a_1}$ and $\mathbf{a_2}$.}
	\label{Ravets:fig:4}
\end{figure}

Using the same nanofabrication techniques described above, micropillar cavities can be used as building blocks to assemble regular arrays of coupled resonators. Figure~\ref{Ravets:fig:4} shows a scanning electron microscope image of such a structure, consisting of overlapping micropillars arranged in a periodic fashion. In this geometry, each heterostructure layer acquires a spatially varying refractive index profile $n(\mathbf{r})$: the refractive index takes the value $n_{\rm layer}$ where material is present, and $n=1$ where material has been etched away. Such microstructure supports a series of electromagnetic modes, which are solution of the three-dimensional Maxwell equation Eq.~\ref{Ravets:eq:Maxwell3D}. Directly solving this equation for large lattices is numerically costly. It is therefore useful to introduce a series of approximations that yield a simplified description. 

\subsubsection{Mapping to a 2D Schr\"odinger equation}

Let us recall the vector identity:
\begin{equation}
\nabla \times \nabla \times \mathbf{E} = \nabla(\nabla \cdot \mathbf{E}) - \nabla^2 \mathbf{E} \;.
\label{Ravets:eq:EfieldIdentity}
\end{equation}
A first simplification is introduced through the so-called ``weak-guidance approximation'', which consists in neglecting the term $\nabla(\nabla \cdot \mathbf{E})$ in the Maxwell equation. This approximation, widely used in fiber optics theory, is valid when the refractive index contrast $\Delta n$ in the medium is small ($\Delta n/n \ll 1$), as is typically the case between the core and the cladding of an optical fiber~\cite{snyder1983}. We note that in micropillar cavities, the index contrast between the semiconductor material and vacuum is larger, so that the approximation is less accurate. Nevertheless, it remains extremely useful, as it enables us to greatly simplify the problem while still capturing the essential physics.

Next, we assume that the modes are predominantly transverse, $|E_z| \ll |E_\perp|$. Within this approximation, the longitudinal variation of the field along the cavity axis $z$ is dominant, such that derivatives of the weak longitudinal component $E_z$ 
with respect to the transverse coordinates can be neglected. The field can then be written as:
\begin{equation}
\mathbf{E}_j(\mathbf{r}) \simeq E_j(\mathbf{r}) \, \mathbf{e_P} \;,
\end{equation}
where $\mathbf{e_P}$ is a polarization direction. This approximation corresponds to a scalar representation in which the polarization degree of freedom is assumed to be decoupled from the orbital (transverse) dynamics. Under these assumptions, the Maxwell equation reduces to the scalar Helmholtz equation:
\begin{equation}
\nabla^2 E(\mathbf{r}) + \left( \frac{\omega}{c} \right)^2 n^2(\mathbf{r}) \, E(\mathbf{r}) = 0 \;.
\end{equation}

In the micropillar geometry, the discrete modes can be described as guided waves propagating along the $z$ axis, and written in the form:
\begin{equation}
E_j(x,y,z) = \psi_j(x,y) \, e^{i \beta_j z} \;,
\end{equation}
where $\beta_j$ is the propagation constant of mode $j$. Substituting into the Helmholtz equation leads to the transverse eigenvalue problem:
\begin{equation}
\nabla_{xy}^2 \psi_j(x,y) + \left[ \left(\frac{\omega}{c}\right)^2 n^2(x,y) - \beta_j^2 \right] \psi_j(x,y) = 0 \;.
\end{equation}
It is then convenient to take the planar cavity as a reference, with the bottom of its dispersion defining the reference energy $\hbar \omega_0$, and the effective refractive index $n_{\text{eff}}$ introduced in Eq.~\ref{Ravets:eq:neff} setting the corresponding reference propagation constant $\beta_0 = \omega_0 n_{\text{eff}} / c$. For a confined mode, the propagation constant can then be expressed as
\begin{equation}
\beta_j = \beta_0 - \delta \beta_j \;,
\end{equation}
where $\delta \beta_j$ encodes a small correction related to the lateral confinement in each mode. Expanding in $\delta \beta_j$ and rewriting the result in energy units, we obtain the following equation for $\psi_j$:
\begin{equation}
\left[ -\frac{\hbar^2}{2 m_{\text{ph}}} \nabla_{xy}^2 + V(x,y) \right] \psi_j(x,y) = \epsilon_j \, \psi_j(x,y) \;,
\end{equation}
with $\epsilon_j = \frac{\hbar c}{n_{\text{eff}}} \, \delta \beta_j$ and $V(x,y)$ an optical confinement potential:
\begin{equation}
V(x,y) = \frac{\hbar \omega_0}{2}\left[1 - \left(\frac{n(x,y)}{n_{\text{eff}}}\right)^2 \right] \;.
\end{equation}
Therefore, the original Maxwell problem maps onto a two-dimensional Schr\"odinger equation describing a particle of mass $m_{\text{ph}}$ moving in a periodic optical potential $V(x,y)$. This effective description provides both the transverse mode profiles $\psi_j(x,y)$ and the discrete eigenenergies $\epsilon_j$, and captures with good accuracy the essential physics of 3D microcavity arrays.

\subsubsection{Linear combination of atomic orbitals}

We now reduce the 2D continuous eigenproblem to a discrete tight-binding model. 
The cavity lattice consists of micropillar resonators centered at positions $(x_j,y_j)$ in the transverse plane. Each pillar supports a localized photonic $s$-orbital $\phi^{(s)}_j(x,y)=\phi^{(s)}(x-x_j,y-y_j)$. By analogy with the Linear Combination of Atomic Orbitals (LCAO) approach~\cite{Ashcroft1976, Kittel2005}, we expand a photonic mode $\psi(x,y)$ as a linear superposition of these orbitals:
\begin{equation}
\psi(x,y)=\sum_j c_j\,\phi^{(s)}_j(x,y) \;,
\end{equation}
where $c_j$ is the complex amplitude on site $j$. Inserting this ansatz into the 2D Schr\"odinger equation yields
\begin{equation}
\sum_j c_j \, \hat{H} \, \phi^{(s)}_j(x,y) = \epsilon \, \sum_j c_j \, \phi^{(s)}_j(x,y) \;,
\end{equation}
with
\begin{equation}
\hat{H}=-\frac{\hbar^2}{2m_{\rm ph}}\nabla_{xy}^2+V(x,y) \;.
\end{equation}
Projecting onto the orbital $\phi^{(s)}_i$ gives:
\begin{equation}
\sum_j H_{ij} c_j = \epsilon \sum_j S_{ij} c_j \;,
\label{Ravets:eq:generalized}
\end{equation}
where we have defined the following coefficients:
\begin{align}
H_{ij} &= \iint dx\,dy\;\phi^{(s)*}_i(x,y)\,\hat{H}\,\phi^{(s)}_j(x,y) 
= \braket{\phi^{(s)}_i|\hat H|\phi^{(s)}_j} \;,\\
S_{ij} &= \iint dx\,dy\;\phi^{(s)*}_i(x,y)\,\phi^{(s)}_j(x,y) 
= \braket{\phi^{(s)}_i|\phi^{(s)}_j} \;.
\end{align}
The coefficients $S_{ij}$ are overlap integrals between neighboring orbitals. Such overlap integrals between orbitals localized on different pillars decay exponentially with distance and are typically orders of magnitude smaller than the Hamiltonian matrix elements $H_{ij}$. Therefore, assuming that $S_{ij} = \delta_{ij}$, Eq.~\ref{Ravets:eq:generalized} reduces to a discrete eigenvalue problem:
\begin{equation}
\sum_j H^{\rm TB}_{ij}\,c_j = \epsilon\, c_i \;,
\end{equation}
where the operator $\hat{H}_{\rm TB}$ has diagonal elements corresponding to onsite energies, $H_{ii}\simeq \epsilon_s$, while the off-diagonal terms correspond to hopping amplitudes $J_{ij}=-H_{ij}$, typically restricted to nearest-neighbor coupling. This establishes the mapping from the continuous photonic problem to a discrete tight-binding Hamiltonian that reads, in bra-ket notation ($\ket{i}\equiv\ket{\phi^{(s)}_i}$):
\begin{equation}
\hat{H}_{\rm TB} = \epsilon_s\sum_i \ket{i}\bra{i} - \sum_{i\neq j} J_{ij}\,\ket{i}\bra{j} \;.
\end{equation}

\subsubsection{Experimental measurement of photonic band structures}

We now turn to the example of the honeycomb lattice shown in Fig.~\ref{Ravets:fig:4}a. 
This lattice contains two sites per unit cell, denoted $A$ and $B$, arranged on a triangular Bravais lattice. 
Within the tight-binding approximation, the real-space Hamiltonian can be written as
\begin{equation}
\hat{H}_{\rm hc} = \epsilon_s \sum_{i} \Big( \ket{i,A}\bra{i,A} + \ket{i,B}\bra{i,B} \Big)
- J \sum_{\langle i,j\rangle} \Big( \ket{i,A}\bra{j,B} + \ket{j,B}\bra{i,A} \Big) \;,
\end{equation}
where $i$ labels the unit cell, $A$ and $B$ denote the sublattice, $J$ is the nearest-neighbor hopping amplitude, and $\langle i,j\rangle$ denotes pairs of nearest neighbors. The first term accounts for onsite energies of the $s$-orbitals, while the second describes hopping between nearest-neighbor sites belonging to opposite sublattices.

\begin{figure}[ht]
    \centering
	\includegraphics[width=0.9\textwidth]{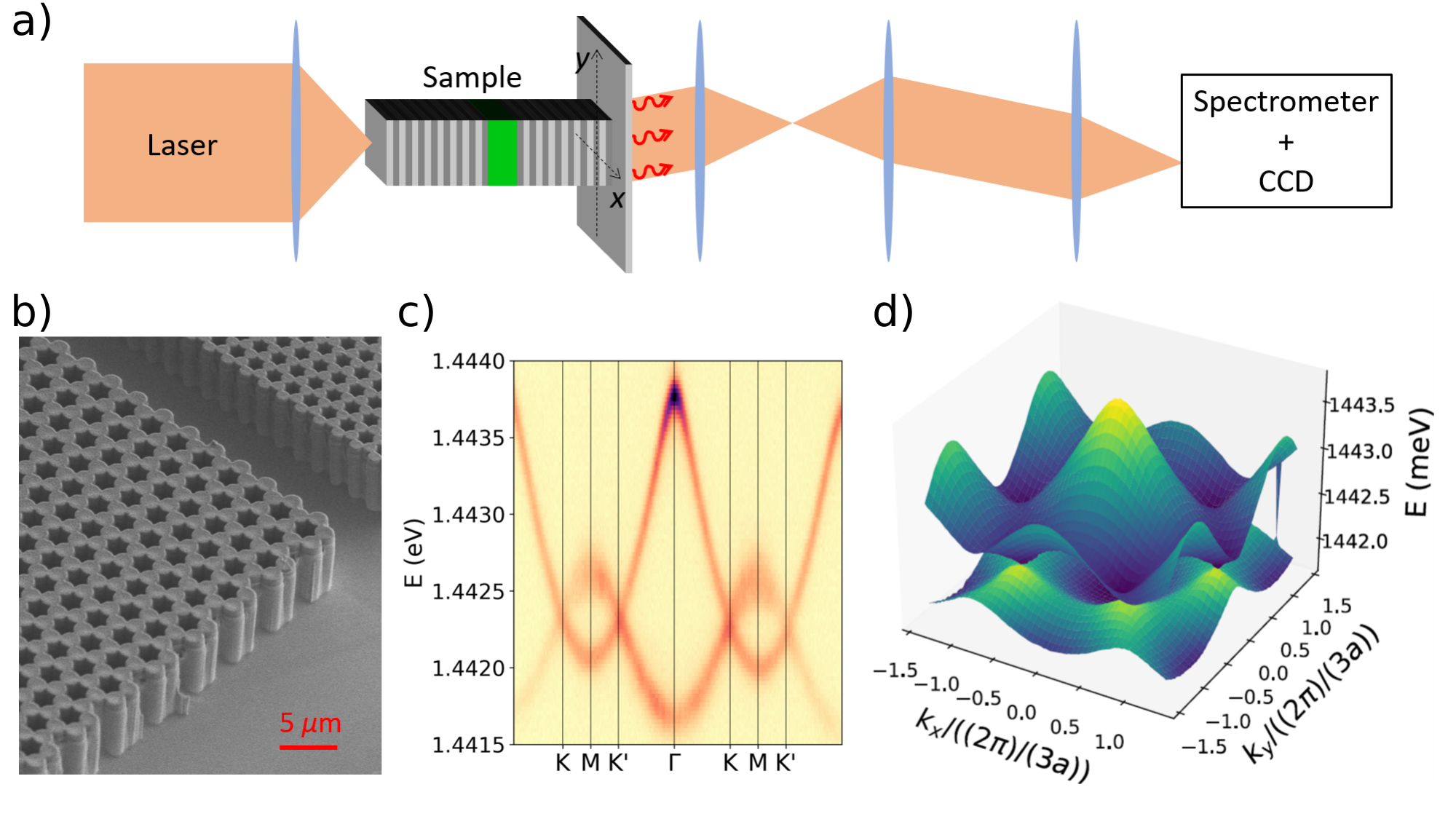}
	\caption{Experimental measurement of photonic band structures. 
	\textbf{(a)} Schematic of the experimental setup: the Fourier-plane emission is imaged onto the entrance slit of a spectrometer. 
	\textbf{(b)} Scanning electron microscope image of the fabricated lattice. 
	\textbf{(c)} Measured emission intensity as a function of $k_x$ for $k_y=0$. 
	\textbf{(d)} By repeating the measurement for multiple values of $k_y$, the full three-dimensional band structure is reconstructed.}
	\label{Ravets:fig:5}
\end{figure}

Using Bloch's theorem, one can write the Bloch Hamiltonian, in the sublattice basis $\{\ket{A,\mathbf{k}},\ket{B,\mathbf{k}}\}$:
\begin{equation}
\hat{H}_{\rm hc}(\mathbf{k}) =
\begin{pmatrix}
\epsilon_s & f(\mathbf{k}) \\
f^*(\mathbf{k}) & \epsilon_s
\end{pmatrix} \;,
\end{equation}
with
\begin{equation}
f(\mathbf{k}) = -J \sum_{j=1}^{3} e^{-i \mathbf{k} \cdot \boldsymbol{\delta}_j} \;,
\end{equation}
where $\boldsymbol{\delta}_j$ are the vectors connecting an $A$ site to its three nearest-neighbor $B$ sites. Diagonalizing $\hat{H}_{\rm hc}(\mathbf{k})$ yields two eigenergies:
\begin{equation}
\epsilon_\pm(\mathbf{k}) = \epsilon_s \pm |f(\mathbf{k})| \;.
\end{equation}
This Hamiltonian supports two dispersive bands, corresponding to bonding and antibonding combinations of the $A$ and $B$ sublattices. 
In reciprocal space, the two bands touch at the six corners of the Brillouin zone, giving rise to Dirac points with linear dispersion, directly analogous to electrons in graphene~\cite{CastroNeto2009}.

Experimentally, the photonic band structure can be directly measured using optical spectroscopy techniques. A schematic of the experimental setup is shown in Fig.~\ref{Ravets:fig:5}a. The sample is maintained at cryogenic temperature ($\sim \qty{4}{\kelvin}$) and excited with a non-resonant laser in the near-infrared ($\lambda \approx \qty{780}{\nano\meter}$). This non-resonant excitation allows populating all the photonic bands, and generates a photoluminescence signal that we then collect in transmission using a confocal microscope configuration. In order to measure the band dispersion, we image the Fourier plane of the microscope objective (Fig.~\ref{Ravets:fig:5}a. We use a set of two lenses aligned in a way that all photons emitted at a given angle (corresponding to a specific in-plane momentum $\mathbf{k_{\parallel}}$) are focused onto a single point at the entrance slit of a spectrometer. The spectrometer then disperses the light in the orthogonal direction, providing simultaneous resolution in both energy and in-plane momentum. The resulting two-dimensional image recorded on the CCD camera directly maps the photonic band structure $E(k_x)$ for a fixed $k_y$, as shown in Fig.~\ref{Ravets:fig:5}c. Scanning the lens in front of the spectrometer, we reconstruct the full three-dimensional band structure $E(k_x,k_y)$ shown in Fig.~\ref{Ravets:fig:5}d.

The experimental dispersion in Fig.~\ref{Ravets:fig:5}c clearly reveals two bands originating from the hybridization of $s$-orbitals on the two sublattices. As expected from the tight-binding Hamiltonian, the bands cross at six Dirac points located at the corners of the Brillouin zone. This observation demonstrates that photons confined in such microstructured lattices can faithfully emulate the physics of electrons in graphene, including the characteristic linear dispersion near the Dirac points. In practice, the measured band structure is not perfectly symmetric, in contrast with the predictions of the simplest tight-binding model. These deviations arise from higher-order effects neglected in the basic description, such as the contribution of higher orbital modes~\cite{Jacqmin2014}. Accounting for these corrections, either by using an extended tight-binding model~\cite{Mangussi2020} or by directly solving the two-dimensional Schrödinger equation, yields a more accurate description of the experimental results.

This example illustrates the power of photonic lattices as emulators of crystalline band structures: complex geometries can be engineered at the micron scale, and their band dispersions can be directly measured using purely optical techniques. 
Beyond dispersion relations, a variety of observables can be accessed in these systems. Recent experiments have demonstrated the possibility of reconstructing the eigenmodes with both amplitude and phase information, together with their eigenenergies~\cite{Gianfrate2020, guillot2025}. This enables a complete characterization of the system, including its underlying quantum geometry~\cite{Bleu2018}.

In summary, we have seen that:
\begin{itemize}
\item Lateral confinement creates discrete photonic orbitals that can be coupled in lattices. 
\item Coupled cavities enable the engineering and direct optical measurement of tailored photonic band structures.
\end{itemize}

\section{Engineering photon-photon interactions: Exciton-polaritons}
\label{Ravets:sec:3}

So far, we have seen that photons can behave as massive particles trapped in engineered potentials. Up to now, we have focused on single particle physics, where the description has remained completely linear. In this section we turn to nonlinear physics, and study how one can use a nonlinear medium to engineer effective photon--photon interactions.

We focus here on the case of a single resonator (a micropillar cavity), restricted to a single mode, the $s$-orbital. Following the discussion in Sect.~\ref{Ravets:sec:2}, the cavity Hamiltonian for a single site writes, in second quantization:
\begin{equation}
\hat H_{\rm cav} = \epsilon_s \, \hat{p}^\dagger \hat{p} \;,
\end{equation}
where $\hat{p}$ annihilates a photon in the $s$-mode. $\hat{H}_{\rm cav}$ corresponds to the Hamiltonian of a harmonic oscillator: its eigenstates form a harmonic energy ladder, with evenly spaced levels. In other words, the energy cost for adding $N$ photons in the cavity is exactly $N \epsilon_s$, with no additional energy required to reach multiple occupancy. The bare cavity photons indeed behave as non-interacting particles.

In order to induce photon-photon interactions, one needs to introduce anharmonicities into this energy ladder. This can be achieved by coupling the harmonic oscillator to a nonlinear system, as we now recall with the Jaynes-Cummings model.

\subsection{Cavity photons coupled to a single two-level system}
\label{Ravets:subsec:3.1}

\begin{figure}[htb]
    \centering
	\includegraphics[width=0.9\textwidth]{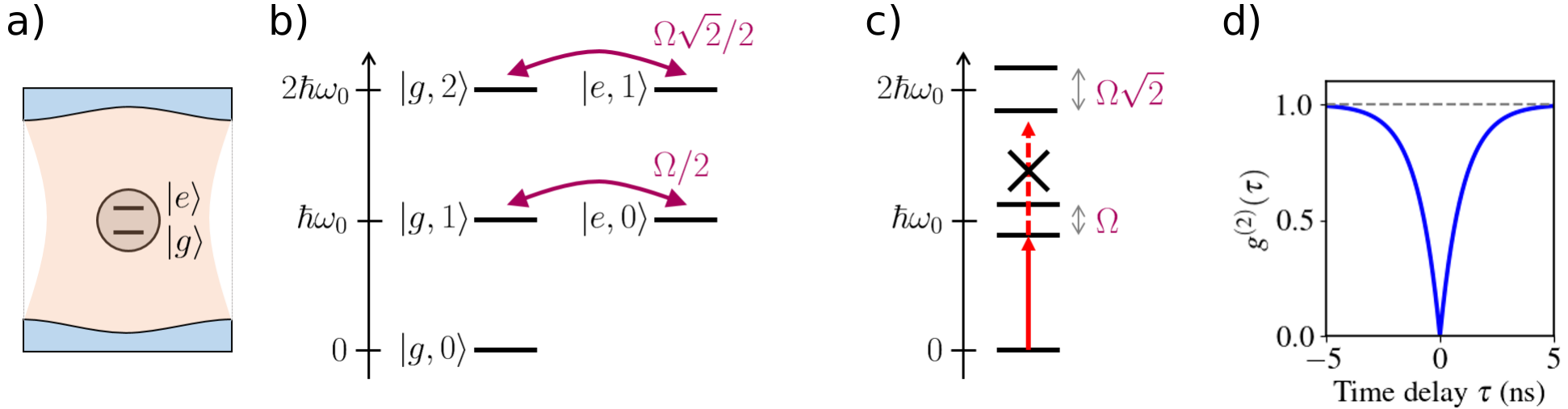}
	\caption{Coupling of cavity photons to a single two-level system. 
	\textbf{(a)} Schematic of the experimental setup. 
	\textbf{(b)} Energy levels in the bare basis, with the coupling terms in the Hamiltonian indicated by magenta arrows. 
	\textbf{(c)} Energy ladder in the normal-mode (dressed-state) basis, which becomes anharmonic as illustrated by the red arrows. 
	\textbf{(d)} This anharmonicity gives rise to the photon blockade effect, resulting in photon antibunching.}
	\label{Ravets:fig:6}
\end{figure}

We consider a single cavity mode resonantly coupled to a two-level system with ground and excited states $\left \{ \ket{g} ; \ket{e} \right \}$ (Fig.~\ref{Ravets:fig:6}a). The coupled system is described by the Jaynes--Cummings Hamiltonian:
\begin{equation}
\hat H_{JC} = \hbar \omega_0 \, \hat{p}^\dagger {p} \; + \; \frac{\hbar \omega_{0}}{2} \, \hat{\sigma}_z \;+\; \hbar \frac{\Omega}{2} \, \big( \hat{p}^\dagger\hat{\sigma}_- + \hat{p} \hat{\sigma}_+ \big) \;, 
\end{equation}
where $\omega_0$ is the cavity frequency, resonant with the atomic transition frequency, $\Omega$ is the vacuum Rabi frequency, and $\hat{\sigma}_\pm$, $\hat{\sigma}_z$ are Pauli operators describing the two-level system. The interaction term couples the states $\ket{g,n}$ and $\ket{e,n-1}$ with a matrix element $\Omega \sqrt{n}/2$ (Fig.~\ref{Ravets:fig:6}b).

Diagonalization of $\hat H_{JC}$ yields ``dressed states'', which are hybrid light–matter excitation, mixing the cavity photon and the atomic excitation. These eigenstates are arranged in doublets exhibiting an energy splitting that scales as $\Omega \sqrt{N}$, where $N$ is the number of excitations in the system. As a result, successive transitions are no longer equally spaced, and the energy ladder of the coupled system is anharmonic (Fig.~\ref{Ravets:fig:6}c). Indeed, by coupling a harmonic oscillator (the cavity mode) with a nonlinear system (the two-level system), one imprints the nonlinearity of the atom onto the photons. 

This anharmonicity is at the origin of the photon blockade effect (Fig.~\ref{Ravets:fig:6}d), introduced in 1997 by A.~Imamoglu and coworkers~\cite{Imamoglu1997}. If a laser is tuned in resonance with the first rung of the ladder (the one-photon excitation), then the laser is detuned from the second rung, which cannot be reached resonantly. Consequently, only a single photon can enter the cavity at a time, effectively realizing a photon-photon interaction. In the output photon statistics, this manifests as photon antibunching, with the zero-delay correlation function
\begin{equation}
g^{(2)}(0) \simeq 0 \;,
\end{equation}
signaling single-photon emission. This effect was first observed using single atoms in high-finesse cavities~\cite{Birnbaum2005}. In solid-state systems, artificial atoms such as self-assembled quantum dots play the role of two-level emitters~\cite{Marzin1994}. When embedded in microcavities, they have been used to implement a solid-state version of the Jaynes--Cummings model in the strong-coupling regime~\cite{Yoshie2004, Reithmaier2004, Najer2019}.

For the purpose of engineering synthetic quantum materials made of solid state arrays of coupled nonlinear cavities, a major limitation arises here. Indeed, in the solid state, two-level systems arising from electronic spatial confinement suffer from strong inhomogeneous broadening~\cite{Marzin1994}, and it is impossible with current technologies to fabricate perfectly identical emitters. This prevents scaling up the Jaynes-Cummings scheme to large arrays in the solid state. As an alternative approach to scalable nonlinear photonic systems, we next consider the case of homogeneous two-dimensional nonlinear media embedded in cavities, forming exciton-polaritons.

\subsection{Quantum well excitons}
\label{Ravets:subsec:3.2}

So far, we have discussed the dielectric properties of semiconductor materials, and their use to realize optical devices. Semiconductor materials also possess rich electronic properties stemming from their band structure. For semiconductors at low temperature, the valence band is filled with electrons while the conduction band remains empty, the two being separated by a finite band gap. The energy gap is typically smaller than in insulators ($E_{\rm gap} \simeq \qty{1.519}{\electronvolt}$ in GaAs), which makes it possible to promote electrons across the gap by optical excitation. Promoting an electron from the valence band to the conduction band leaves a hole behind, which can be treated as a positively charged particle. The resulting electron–hole pair interacts via Coulomb attraction and forms bound states called excitons~\cite{Bastard1988}. The exciton energy dispersion reads:
\begin{equation}
E_X({n,\mathbf{K}}) = E_{\rm Gap} - \frac{E_{\rm B}}{n^2} + \frac{\hbar^2 \mathbf{K}^2}{2 M_X} \;,
\label{Ravets:eq:ExcitonDispersion}
\end{equation}
where $E_{\rm B}$ is the exciton binding energy, $n$ is the principal quantum number (typically $n=1$ in GaAs), $\mathbf{K}$ is the center of mass momentum, and $M_X$ is the exciton mass. We note that a typical order of magnitude for the exciton mass is $M_X \sim m_{e^-}$, typically five orders of magnitude larger than the effective photon mass in the cavity.

\begin{figure}[ht]
    \centering
	\includegraphics[width=0.9\textwidth]{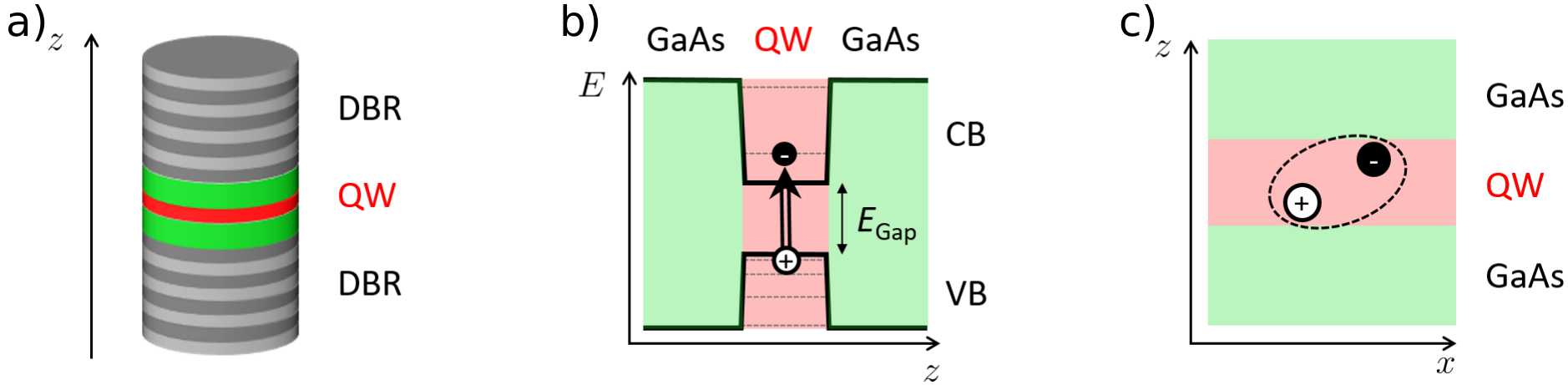}
	\caption{Quantum well excitons coupled to cavity photons.
	\textbf{(a)} Schematic of a micropillar cavity embedding a semiconductor quantum well at its center.
	\textbf{(b)} Bandgap engineering creates a confining potential for charge carriers. Under optical excitation (double arrow), an electron is promoted from the valence band (VB) to the conduction band (CB), leaving behind a hole.
	\textbf{(c)} The Coulomb attraction between the electron and the hole binds them into an exciton.}
	\label{Ravets:fig:7}
\end{figure}

Another interesting feature of semiconductor materials is that the band gap depends on the alloy composition, and can be tuned by incorporating different atomic species. For example, adding a few percent of indium to GaAs, yields $\rm{In}_p\rm{Ga}_{1-p}\rm{As}$ alloys, which have a reduced bandgap energy. Therefore, the ability to combine and stack different materials enables bandgap engineering, opening the possibility to design potential barriers for charge carriers and to confine excitons (Fig.~\ref{Ravets:fig:7}b). A semiconductor quantum well is a thin layer (typically a few nanometers) of smaller-gap material embedded in a larger-gap material, confining electrons and holes along the growth axis. The confinement adds a static trapping potential along one direction, while the electron–hole pair remains free to move in the plane. The exciton dispersion remains similar to Eq.~\ref{Ravets:eq:ExcitonDispersion}, with slightly renormalized energies to account for the confinement, and a ground-state binding energy approximately four times larger than in bulk. Thanks to their enhanced binding energy (smaller Bohr radius), 2D excitons couple efficiently to light. Moreover, quantum wells can be positioned precisely at the antinodes of a cavity mode (Fig.~\ref{Ravets:fig:7}c), optimizing the strength of light–matter coupling.

Excitons are composite bosons at low density and can be described by bosonic operators. Because the exciton mass is heavy compared to the photon mass, a collective excitonic excitation can follow the cavity mode profile and couple efficiently to the photon. We thus define the operator $\hat{x}$, which creates an exciton in the same spatial mode as the photon. Being composed of charged particles, excitons interact through Coulomb interactions, giving rise to a nonlinear term in the Hamiltonian due to the exchange interaction between their fermionic constituents~\cite{Ciuti1998, Tassone1999, Carusotto2013}:
\begin{equation}
\hat H_X \;=\; \epsilon _X \, \hat{x}^\dagger \hat{x} \;+\; \frac{U_X}{2}\, \hat{x}^\dagger \hat{x}^\dagger \hat{x} \hat{x} \;,
\end{equation}
where $U_X$ is the exciton–exciton interaction strength. This nonlinear term manifests as an anharmonic energy ladder for the bosonic excitons: creating two excitons necessitates an energy $2 \epsilon_X +U_X$, where $U_X$ represents the extra interaction energy. 

We emphasize that excitons behave differently than the two-level systems featured in the Jaynes–Cummings model. Nevertheless, we show below that their intrinsic nonlinearity can still be transferred to photons \textit{via} the light–matter coupling in micropillar cavities.

\subsection{Strong coupling regime: exciton polaritons}
\label{Ravets:subsec:3.3}

\begin{figure}[ht]
    \centering
	\includegraphics[width=0.9\textwidth]{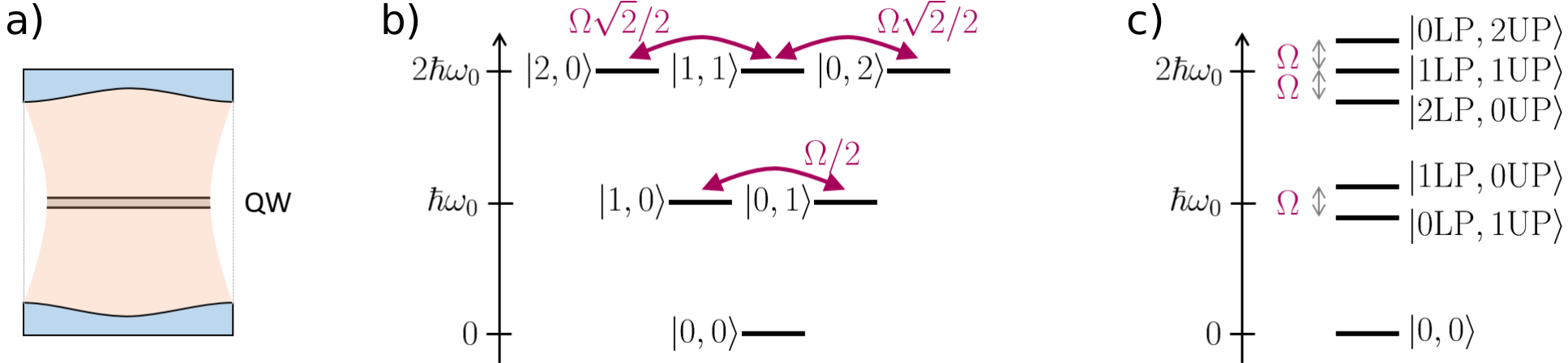}
	\caption{Exciton-polaritons in a single mode cavity.
	\textbf{(a)} Schematic of a single mode cavity embedding a quantum well at its center.
	\textbf{(b)} Linear Hamiltonian energy levels in the bare basis, with the coupling terms indicated by magenta arrows.
	\textbf{(c)} Energy ladder in the polariton basis, for the linear Hamiltonian.}
	\label{Ravets:fig:8}
\end{figure}

We now consider a micropillar cavity embedding a semiconductor quantum well at its center (Fig.~\ref{Ravets:fig:8}a. We assume that the cavity resonance is tuned to the exciton transition energy ($\epsilon_X = \epsilon_s$). The total Hamiltonian of the coupled system can be written as:
\begin{equation}
\hat{H}_{\rm tot} = \epsilon_s \, (\hat{p}^\dagger \hat{p} + \hat{x}^\dagger \hat{x}) 
+ \hbar \frac{\Omega}{2} \, (\hat{p}^\dagger \hat{x} + \hat{x}^\dagger \hat{p})
+ \frac{U_X}{2} \, \hat{x}^\dagger \hat{x}^\dagger \hat{x} \hat{x} \;,
\end{equation}
where $\Omega$ denotes the vacuum Rabi splitting arising from the dipole interaction between the quantum-well exciton and the confined cavity mode. Its magnitude is fixed by the exciton dipole moment and by the strength of the vacuum electric field of the cavity mode at the position of the quantum well.

We first focus on the linear part $\hat{H}_{\rm lin}$ of the Hamiltonian ($U_X = 0$, Fig.~\ref{Ravets:fig:8}b. Because of light–matter coupling, the eigenmodes are expected to be superpositions of the bare exciton and photon modes. Introducing the symmetric and antisymmetric combinations:
\begin{align}
\hat{l} &= \frac{1}{\sqrt{2}} \, (\hat{p} - \hat{x}) \;, \\
\hat{u} &= \frac{1}{\sqrt{2}} \, (\hat{p} + \hat{x}) \;,
\end{align}
and rewriting the Hamiltonian in this basis, we obtain the diagonal form:
\begin{equation}
\hat{H}_{\rm lin} = \left( \epsilon_s - \hbar \frac{\Omega}{2} \right) \, \hat{l}^\dagger \hat{l} \;+\; \left( \epsilon_s + \hbar \frac{\Omega}{2} \right) \hat{u}^\dagger \hat{u} \;.
\end{equation}

This shows that the (single particle) eigenstates of the linear Hamiltonian are coherent superpositions of photons and excitons, known as polaritons:
\begin{align}
\ket{1_l,0_u} &= \hat{l}^\dagger \ket{0,0} = \frac{1}{\sqrt{2}} (\ket{1_p , 0} - \ket{0, 1_x}) \;, \\
\ket{0_l,1_u} &= \hat{u}^\dagger \ket{0,0} = \frac{1}{\sqrt{2}} (\ket{1_p , 0} + \ket{0, 1_x}) \;.
\end{align}
Physically, the excitation undergoes Rabi oscillations between its photonic and excitonic components, spending part of its time as light and part as matter. The lower energy mode ($\hat{l}$) is the lower polariton, while the higher energy mode ($\hat{u}$) is the upper polariton.

We note that here, although we coupled light and matter excitations, the spectrum of $\hat{H}_{\rm lin}$ remains harmonic. This is expected, since both excitons and photons behave as harmonic oscillators, in stark contrast with the Jaynes–Cummings model where the matter degree of freedom is a nonlinear two-level system. To restore anharmonicity in the polariton ladder, we now include exciton–exciton interactions through the nonlinear term.

Assuming the Rabi coupling is large compared with other relevant energy scales, the upper and lower polariton branches are well separated, and we can focus on the lower polaritons. At low density, the interaction can be treated perturbatively, leading to the lower polariton Hamiltonian:
\begin{equation}
\hat{H}_{\rm LP} = \epsilon_{\rm LP} \, \hat{l}^\dagger \hat{l} 
\;+\; \frac{U_{\rm LP}}{2} \, \hat{l}^\dagger \hat{l}^\dagger \hat{l} \hat{l} \;,
\end{equation}
where $\epsilon_{\rm LP} = \epsilon_s -\hbar \Omega/2$, and the lower-polariton interaction constant is $U_{\rm LP} = U_X /4$, reflecting the fact that polaritons are only half-matter. Injecting two polaritons into the cavity thus costs an additional energy $U_{\rm LP}$, restoring an anharmonic energy ladder.

This effect is the polariton analogue of the Jaynes–Cummings model, and can give rise to polariton blockade~\cite{Verger2006}, where the presence of one polariton inhibits the injection of a second. The drawback compared to Jaynes-Cummings systems is that the nonlinearity $U_{\rm LP}$ is typically much weaker than in the two-level system case (see Sect.~\ref{Ravets:sec:4}). Nevertheless, the approach is highly scalable, since homogeneous quantum wells can be fabricated and patterned into arbitrary cavity arrays with minor inhomogeneous broadening, as we will see next.

\subsection{Driven–dissipative Bose–Hubbard model}
\label{Ravets:subsec:3.4}

Combining photon hopping between neighboring resonators with on-site polariton–polariton interactions yields an effective lattice Hamiltonian of the form:
\begin{equation}
\hat{H}_{\rm LP} =
\sum_n 
\epsilon_{\rm LP} \, \hat{l}_n^\dagger \hat{l}_n
+ \frac{U_{\rm LP}}{2} \, \hat{l}_n^\dagger \hat{l}_n^\dagger \hat{l}_n \hat{l}_n
- t \sum_{\langle n,m \rangle} \big( \hat{l}_n^\dagger \hat{l}_m + \hat{l}_m^\dagger \hat{l}_n \big) \;.
\end{equation}
This Hamiltonian is formally identical to the celebrated Bose–Hubbard model, where the operators $\hat{l}_n$ now describe lower polariton modes localized on each site of the lattice.

It is important to stress that photonic are intrinsically open: photons have a finite lifetime and continuously leak out of the cavity at a rate $\gamma$. As a result, the system is intrinsically open and must be driven to maintain a steady population of polaritons. For example, this can be achieved by coupling the cavity to a coherent pumping field, with frequency $\omega_{\rm P}$ and complex amplitude $F_n$ on site $n$~\cite{Carusotto2025}. Including this drive, the total Hamiltonian reads:
\begin{equation}
\hat{H}_{\rm tot} =
\hat{H}_{\rm LP}
+ i \hbar \sum_n \left( F_n e^{-i \omega_{\rm P} t } \, \hat{l}_n^\dagger
- F_n^* e^{i \omega_{\rm P} t } \, \hat{l}_n \right) \;.
\end{equation}
The open nature of the system is fully captured by a master equation for the density matrix $\hat{\rho}$,
\begin{equation}
\frac{d \hat{\rho}}{dt}
= - \frac{i}{\hbar} \big[\hat{H}_{\rm tot}, \hat{\rho}\big] 
+ \gamma \sum_n \mathcal{L}[\hat{l}_n] \hat{\rho} \;,
\label{Ravets:eq:mastereq}
\end{equation}
with $\mathcal{L}[\hat{l}] \hat{\rho}
= \hat{l} \hat{\rho} \hat{l}^\dagger
- \tfrac{1}{2} \big\{ \hat{l}^\dagger \hat{l}, \hat{\rho} \big\} $ the Lindblad dissipator describing photon loss.

This is the driven–dissipative Bose–Hubbard model, which constitutes the general theoretical framework for polariton lattices. It highlights both the similarities with typical many-body models and the essential differences due to drive and dissipation. This framework provides the starting point for exploring a wide variety of phenomena, as we review in the next Section:
\begin{itemize}
\item Coupling cavity photons to electronic excitations (quantum well excitons) gives rise to hybrid light–matter quasiparticles called polaritons. 
\item Owing to their excitonic component, polaritons interact with each other, which manifests as a nonlinear term in the Hamiltonian. 
\item Arrays of coupled cavities hosting interacting polaritons realize a driven–dissipative version of the Bose–Hubbard model.
\end{itemize}

\section{Synthetic matter in the exciton-polariton platform: an overview}
\label{Ravets:sec:4}

We have seen that polariton lattices can be used to implement driven-dissipative Bose-Hubbard physics, providing a highly versatile platform for simulating synthetic matter. In this section, we review some realizations of synthetic materials in the exciton-polariton platform, and highlight current research directions, covering the linear regime, the mean-field regime, and perspectives towards the quantum regime.

\subsection{Hamiltonian engineering in the linear regime}
\label{Ravets:subsec:4.1}

The mapping of the two-dimensional lattice system to tight-binding Hamiltonians discussed in Sect.~\ref{Ravets:sec:2} enables implementing a wide variety of models and exploring various aspects of lattice physics, as we briefly review below.

\subsubsection{Lattice physics in engineered potentials}

By leveraging the versatility of nanofabrication tools, the lattice geometry and therefore the corresponding tight-binding Hamiltonian can be tailored at will: the pillar diameter sets the onsite energy, the interpillar distance controls the hopping amplitude $J$, and the design of the unit cell enables engineering the band structure. This flexibility has opened the way to exploring a broad range of physical phenomena:
\begin{itemize}
	\item \textbf{\emph{Dirac points and semimetals.}} Honeycomb lattices of micropillars (see Sect.~\ref{Ravets:sec:2}) have enabled emulating graphene and studying the physics of two-dimensional semimetals characterized by band-touching points in the Brillouin zone~\cite{Jacqmin2014, Klembt2018, Milicevic2019, Real2020, Jamadi2020}. We point out that lattice engineering provides access to regimes that are inaccessible in real materials: for example introducing controlled strain in the hopping parameters has led to the observation of new types of Dirac cones~\cite{Milicevic2019}.
	\item \textbf{\emph{Flat bands.}} Some studies have focused on realizing flatbands, using Lieb lattices~\cite{Baboux2016, Whittaker2018, Goblot2019, Harder2020} or Kagome lattices~\cite{Gulevich2016, Harder2021}. Owing to their nearly dispersionless character, flatbands strongly enhance the role of interactions, providing a promising platform to explore the interplay between flatband physics and nonlinearities.
	\item \textbf{\emph{Topological physics.}} Polaritonic lattices have been used to explore concepts related to topology and geometry in condensed matter. First explorations were performed in one-dimensional chains where topological edge states were observed~\cite{StJean2017, StJean2021}. In two dimensions, a polariton topological insulator has been proposed~\cite{Karzig2015, Nalitov2015} and then realized experimentally~\cite{Klembt2018}. Artificial gauge fields for polaritons are discussed in more detail in the next paragraph.
	\item \textbf{\emph{Quasi-periodic potentials.}} Going beyond periodic lattices, quasi-periodic structures have been realized, exhibiting fractal energy spectra, and critical wavefunction behavior with intriguing localization properties~\cite{Tanese2014, Baboux2017, Goblot2020}.
	\item \textbf{\emph{Non-hermitian physics.}} Across all these realizations, the intrinsically driven-dissipative nature of polaritons provides a unique opportunity to investigate the role of losses in band structures. This has revealed a rich non-Hermitian phenomenology, including exceptional points, Fermi arcs, and non-orthogonal eigenstates~\cite{Gao2015, Pickup2020, Su2021, Hu2023}.
\end{itemize}

\subsubsection{Artificial gauge fields for polaritons and polariton topological insulators}

So far, we have mainly considered a scalar description of the photon field (Sect.~\ref{Ravets:sec:2}), and the consequences of its hybridization with excitons on polariton–polariton interactions (Sect.~\ref{Ravets:sec:3}). The physics of polaritons, however, is even richer: owing to their mixed light–matter nature, they possess unique polarization properties inherited from the photonic part, while their excitonic component makes them sensitive to external electric and magnetic fields. These additional ingredients open more possibilities for engineering synthetic matter with polaritons. As a striking example, polaritons can be made to behave analogously to charged particles in a magnetic field. Several complementary strategies have been explored, as summarized below:
\begin{itemize}
\item \textbf{\emph{Spin–orbit coupling.}} Because of the vectorial nature of Maxwell’s equations, Bragg microcavities support two polarization modes (TE and TM): the full description therefore requires a spinor field. In practice, the effective photon masses for the two polarizations differ, leading to a TE–TM energy splitting that increases with the in-plane wavevector $\mathbf{k}$. In addition, the polarization axes rotate with $\mathbf{k}$, so that the photon pseudospin (polarization state) becomes coupled to its momentum~\cite{kavokin2005}. This effective spin–orbit coupling has been observed in several contexts, from the optical spin Hall effect in planar cavities~\cite{Leyder2007} to a Dresselhaus-type interaction in honeycomb lattices~\cite{Whittaker2021}.
\item \textbf{\emph{Topological insulators.}} Polaritons are also sensitive to magnetic fields, which allows the introduction of a Zeeman splitting between circularly polarized excitons, effectively breaking time-reversal symmetry. Combining spin–orbit coupling with Zeeman splitting near band-touching points opens topological gaps, giving rise to Chern insulators~\cite{Karzig2015, Nalitov2015, Klembt2018}. Such bands support robust chiral edge states and bear close analogies to the physics of electronic quantum Hall systems.
\item \textbf{\emph{Artificial gauge potentials.}} Another proposed approach exploits the excitonic component of polaritons, which provides sensitivity to applied electric and magnetic fields. By applying a carefully designed combination of static electric and magnetic fields, an electrically tunable artificial gauge potential for polaritons has been realized in planar cavities~\cite{Lim2017}. Extending this strategy to lattice geometries could enable the realization of Harper–Hofstadter models~\cite{Harper1955, Hofstadter1976} featuring a non-zero synthetic magnetic flux per plaquette, enabling the exploration of topological phenomena.
\item \textbf{\emph{Landau levels for photons.}} Strong synthetic magnetic fields have been engineered in honeycomb lattices by imposing a uniaxial gradient in the hopping amplitudes. This approach gives rise to the formation of Landau levels at the Dirac points, as directly observed in polariton lattices~\cite{Jamadi2020}.
\end{itemize}

Overall, the ability to combine lattice geometry, synthetic gauge fields, spin–orbit coupling, and Zeeman splitting enables engineering a wide range of exotic band structures and topological phases. The richness of accessible bands offers an exciting playground for both fundamental studies and potential applications in topological photonics, even at the level of single-particle physics~\cite{Ozawa2019}. Crucially, topology is encoded in the system's eigenstates, not just in the band structure. In this context, the polariton platform is particularly advantageous, as optical techniques allow accessing observables that are otherwise very difficult to measure. For example, new spectroscopic methods have been developed to probe the full quantum geometry of eigenstates in both planar cavities~\cite{Bleu2018, Gianfrate2020} and lattice systems~\cite{guillot2025}. These tools are essential for exploring topological properties and hold promise for going beyond Chern physics, potentially enabling future investigations of multi-gap topology, Euler class, and non-Abelian effects~\cite{Bouhon2020, Bouhon2020b, Slager2024, Finck2025}.

\subsection{Nonlinear optics in the mean-field regime} 
\label{Ravets:subsec:4.2}

Focusing now on the nonlinear regime, the master equation introduced in Eq.~\ref{Ravets:eq:mastereq} for the case of coherent resonant pumping provides a full quantum description of the driven-dissipative Bose--Hubbard model. In many situations of interest, the cavities support a large intracavity photon population, and quantum fluctuations play only a secondary role. In this mean-field limit, the system can be described classically in terms of coherent fields, capturing the essential nonlinear driven-dissipative dynamics~\cite{Carusotto2013}. We review below the cases of resonant and non-resonant pumping. 

\subsubsection{Resonant pumping}

A widely used configuration employs a resonant laser to inject polaritons at a chosen energy. As the pump power increases, the polariton population in the driven state grows large enough to enter the mean-field regime. The mean-field approximation consists in taking the expectation values of the operators, $\psi_j(t) = \langle l_j(t) \rangle$, and neglecting quantum correlations by assuming:
\begin{equation}
\langle l_j^\dagger l_j l_j \rangle \approx |\psi_j|^2 \psi_j \;.
\end{equation}
This assumption, valid for large photon numbers and small relative fluctuations, leads to a set of nonlinear equations:
\begin{equation}
i \hbar \frac{d}{dt} \psi_j =
\left( \epsilon_s - i \hbar \frac{\gamma}{2} \right)\psi_j
+ U_{\rm LP} |\psi_j|^2 \psi_j
- J \sum_{k \in \langle j \rangle} \psi_k
+ \hbar F_j \, e^{- i \omega_{\rm P} t} \;.
\label{Ravets:eq:ddGPE}
\end{equation}
Equation~\ref{Ravets:eq:ddGPE} is a generalized Gross–Pitaevskii equation that includes coherent driving and photon losses. It is commonly referred to as the driven-dissipative Gross–Pitaevskii equation (dd-GPE)~\cite{Carusotto2013}. As a consequence of the Kerr interaction term in the equation, it supports a wealth of nonlinear phenomena, which we summarize below: 
\begin{itemize}
\item \textbf{\emph{Optical bistability.}}
The Kerr term $U|\psi_j|^2$ renders the steady-state solution nonlinear, potentially allowing multiple coexisting states for the same drive frequency and amplitude. Such multistable behavior gives rise to hysteresis cycles when scanning the pump parameters, as was first observed in planar cavities~\cite{Baas2004}, and later in confined microcavity modes~\cite{Paraiso2010} and lattices~\cite{Goblot2019, Pernet2022}. We note that multistability carries the signature of nonequilibrium critical phenomena and dissipative phase transitions~\cite{Casteels2016, Rodriguez2017, Fink2018}. 
\item \textbf{\emph{Optical parametric oscillation.}} The Kerr nonlinearity also enables coherent four-wave mixing, where two pump polaritons convert into a signal and idler pair. Above a threshold, this process becomes parametrically unstable, breaking the $U(1)$ phase symmetry between signal and idler, and leading to macroscopic occupation of these modes. This optical parametric oscillation (OPO) regime was first observed in planar cavities~\cite{Savvidis2000, Stevenson2000, Baumberg2000} and more recently in coupled micropillar dimers~\cite{Zambon2020}. 
\item \textbf{\emph{Polariton hydrodynamics and superfluidity.}} By injecting a planar cavity with a resonant pump at a finite in-plane wavevector, one can generate a flow of polaritons. Linearizing the dd-GPE around the steady-state $\psi_0$:
\begin{equation}
\psi(\mathbf r,t)=\psi_0(\mathbf r) \, e^{-i\omega_P t} + \delta\psi(\mathbf r,t) \;,
\end{equation}
one obtains the Bogoliubov spectrum of excitations. Interactions modify the dispersion, and may produce at some pump power and for small wavevectors, a phonon-like linear branch. According to Landau's criterion, if this branch is steeper than the flow velocity, scattering from defects is suppressed, resulting in frictionless superfluid flow~\cite{Carusotto2004, Carusotto2013}. This has been experimentally observed in GaAs microcavities at $\qty{4}{\kelvin}$~\cite{Amo2009}, which opened the field of polariton hydrodynamics. Soon after, the field was further enriched by experiments demonstrating room-temperature superfluidity, dark and bright solitons, quantized vortices, and turbulence in polariton fluids~\cite{Nardin2011, Amo2011, Sanvitto2011, Lerario2017, Panico2023}.
\item \textbf{\emph{Nonlinearities and topology.}} Another exciting direction is the exploration of the interplay between nonlinearity and topology~\cite{Smirnova2020}. For example, in a one-dimensional polariton SSH lattice, Kerr interactions have allowed the bulk excitation of topological solitons localized on a single pillar, effectively generating an all-optical topological interface for Bogoliubov excitations~\cite{Pernet2022}. Also, spatiotemporal modulation of the Kerr interaction in a one-dimensional chain has been proposed to realize topological pumping of Bogoliubov modes, effectively mapping the system onto a 2D Harper-Hofstadter model~\cite{Ravets2025}. These developments, which connect topological photonics with nonlinear wave phenomena, open exciting perspectives for controlling light through the combined effects of topology and interactions at the mean-field level.
\end{itemize}

Under resonant pumping, the dd-GPE thus provides a unifying framework for describing a broad range of nonlinear phenomena. As we see below, another pumping scheme, namely the non-resonant pumping scheme, further enriches this picture.

\subsubsection{Non-resonant (incoherent) pumping}

The non-resonant pumping scheme uses a laser tuned well above the exciton resonance (or band gap energy), creating high-energy electron–hole pairs that relax and populate a reservoir of excitons. The reservoir can either (i) decay non-radiatively, or (ii) feed polariton modes \textit{via} stimulated scattering. A widely used mean-field description of this situation couples the driven–dissipative Gross–Pitaevskii equation to a rate equation for the (classical) exciton reservoir density~\cite{Porras2002, Wouters2007}:
\begin{align}
i \hbar\frac{d}{dt}\psi_j &= \Big[ \omega_0 - i\hbar\frac{\gamma}{2} + U\,|\psi_j(t)|^2 + 2 g_R\, n_{R,j} + \frac{i \hbar}{2}R\,n_{R,j} \Big] \, \psi_j - J\sum_{k\in\langle j\rangle}\psi_k \;,
\label{Ravets:eq:ddGPE_reservoir}
\\
\frac{d}{dt} n_{R,j} &= P_j - \big(\gamma_R + R\,|\psi_j|^2\big)\,n_{R,j} \;,
\label{Ravets:eq:reservoir_rate}
\end{align}
where $n_{R,j}$ is the density of reservoir excitons, $\gamma_R$ is the reservoir decay rate, $R$ is the stimulated scattering rate, $g_R$ is the mean-field energy shift induced by the reservoir, and $P_j$ is the (spatially dependent) non-resonant pump rate injecting carriers into the reservoir. The term $R\,n_{R,j}$ in Eq.~\ref{Ravets:eq:ddGPE_reservoir} provides a gain contribution, while the term $R|\psi_j|^2 n_{R,j}$ in Eq.~\ref{Ravets:eq:reservoir_rate} accounts for reservoir depletion through stimulated scattering. These coupled equations capture the essential physics of incoherently pumped polaritons in semiconductor microcavities, closely related to laser physics, and capture the physics of polariton condensation as well as exotic states such as polariton supersolids:
\begin{itemize}
\item \textbf{\emph{Driven-dissipative condensates.}} Using incoherent pumping, condensation of polaritons has been demonstrated~\cite{Kasprzak2006, Balili2007, Carusotto2013, Bloch2022}. To gain intuition onto this physics, we first consider the case of a single mode, where the condensate is initially empty ($\psi(0)=0$). In the absence of scattering, the reservoir density then relaxes to $n_R^{(0)} = P/\gamma_R$. Stimulated scattering in Eq.~\ref{Ravets:eq:ddGPE_reservoir} provides a gain $\propto R\,n_R$ that compensates the loss term $\propto \gamma$. The condition $R\,n_R^{(0)} \gtrsim \gamma$ thus defines a condensation threshold $P_\mathrm{th} = \gamma \gamma_R /R$, above which the condensate grows, and macroscopic occupation of a state is reached. The onset of a macroscopically populated condensate is associated with spontaneous breaking of the global $U(1)$ phase symmetry: the condensate picks up a phase, leading to the emergence of long-range coherence~\cite{Richard2005, Kasprzak2006, Balili2007, Carusotto2013, Bloch2022}. We note that because the system is inherently driven and lossy, the condensate is open, leading to a rich phenomenology such as modulational instabilities~\cite{Baboux18} or negative mass condensates~\cite{Baboux18, Wurdack2023} for instance. Furthermore, recent studies have highlighted that the dynamics of the condensate phase can be mapped, under some approximations, onto a stochastic equation belonging to the Kardar–Parisi–Zhang (KPZ) universality class~\cite{Altman2015, He2015, Ji2015}. As a consequence, phase correlations and first-order correlation functions exhibit stretched-exponential scalings with universal exponents belonging to the KPZ universality class, in stark contrast with the behavior of closed condensates~\cite{Hadzibabic2006}. This behavior has been recently observed both in one-dimensional and two-dimensional polariton condensates~\cite{Fontaine2022, widmann2025}.
\item \textbf{\emph{Polariton supersolids.}} The nonlinear interactions between polaritons in the condensate can trigger secondary instabilities at higher pump powers, leading to new thresholds beyond the initial $U(1)$ symmetry-breaking. After the first condensation threshold, where the condensate phase is spontaneously chosen, a secondary instability may emerge in carefully designed systems. This instability leads to the breaking of translational symmetry and the appearance of density-modulated condensate states. Such states, recently observed in semiconductor microcavities, have been identified as polariton supersolids~\cite{muszynski2024, Trypogeorgos2025, Nigro2025, kozhevin2025}. These discoveries open new opportunities to explore new exotic phases of driven–dissipative systems, where condensation and spatial order coexist.
\end{itemize}

In summary, the mean-field approximation has proven to be a remarkably powerful framework for polariton systems, capturing a wide range of phenomena such as optical bistability, superfluidity, nonequilibrium condensation, and even supersolidity. This versatility has established the field of quantum fluids of light~\cite{Carusotto2013}, which offers entirely new possibilities rooted in driven–dissipative nature of system~\cite{Carusotto2025}. In the next section, we move beyond the mean-field description to explore the genuinely quantum regime, where fluctuations and correlations play a central role.

\subsection{Towards polariton quantum matter}

As discussed in Sect.~\ref{Ravets:sec:3}, the hybrid nature of polaritons, combining light effective mass with exciton-mediated interactions, offers a unique platform for engineering strong photon–photon interactions. This opens the way to quantum many-body phases of light, where individual photons display non-classical correlations and collective phenomena. Reaching this regime remains an ambitious goal, promising access to novel states of strongly correlated photonic matter.

\subsubsection{Single-site physics and the quest for polariton blockade}

At the single site level, the hallmark of strong photon--photon interactions is the so-called polariton blockade, where the presence of a single photon in a cavity prevents the entry of a second one due to nonlinear energy shifts~\cite{Verger2006}. The relevant figure of merit is the ratio $U_{\rm LP}/\gamma$ between the interaction energy shift $U_{\rm LP}$ and the polariton linewidth $\gamma$. In present-day semiconductor microcavities, $U_{\rm LP}$ is relatively weak, leading to typical values $U_{\rm LP}/\gamma \sim 0.1$. This places current polariton systems in the regime of weak blockade, where deviations from Poissonian photon statistics are weak.

Nevertheless, first experimental signatures of polariton blockade have been reported, with experiments reporting values of the second-order correlation function $g^{(2)}(0) \simeq 0.95$~\cite{Delteil2019, MunozMatutano2019}. These results mark an important milestone: they demonstrate that even in the weakly interacting regime, polaritons can generate non-classical light. Apart from photon antibunching, other quantum optical effects have been realized in this intermediate regime, such as intensity squeezing~\cite{Boulier2014}, single photon routing~\cite{Cuevas2018, SuarezForero2020}, few-photon nonlinear phase shifts~\cite{Kuriakose2022} and polariton radiative cascades~\cite{Scarpelli2024}. Together, these experiments illustrate that polaritons, despite their weak nonlinearities, already provide access to rich quantum optical physics.

Considerable effort is now being devoted to increasing the effective strength of polariton interactions. Several strategies are being explored. One avenue is to exploit resonant enhancement mechanisms such as polaritonic Feshbach resonances, where tuning close to biexciton resonances amplifies the effective nonlinearity~\cite{Takemura2014, Scarpelli2024}. Another approach is to engineer excitons with intrinsically stronger interactions, for example by using dipolar excitons with static dipole moments~\cite{Cristofolini2012, Rosenberg2018, Togan2018}, or Rydberg excitons with extremely large principal quantum numbers~\cite{Kazimierczuk2014}. A further direction relies on hybridizing excitons with electronic environments, leading to new quasiparticles such as polaron-polaritons in atomically thin semiconductors~\cite{Sidler2017, Tan2020}, or polaritons coupled to correlated electron states in the quantum Hall regime~\cite{Ravets2018, Knueppel2019}. All these efforts must be accompanied by parallel progress in reducing cavity losses, since both larger $U_{\rm LP}$ and smaller $\gamma$ are needed to improve the ratio $U_{\rm LP}/\gamma$ and approach the strong blockade regime.

\subsubsection{From single sites to lattices: synthetic quantum polariton matter}

Extending polariton physics from single sites to arrays of coupled cavities or micropillars opens the door to collective quantum phenomena and many-body physics. In particular, coupled cavity arrays with strong interactions realize a driven-dissipative version of the Bose–Hubbard model, providing an exciting platform for engineering synthetic quantum materials, as we review below.
\begin{itemize}
\item \textbf{\emph{Photon fermionization.}} One striking prediction is the fermionization of photons in one-dimensional arrays: strong on-site interactions can lead photons to behave as hard-core bosons, mimicking the physics of a Tonks-Girardeau gas~\cite{Carusotto2009, Carusotto2013}. In this regime, local photon antibunching coexists with enhanced nonlocal cross-correlations, offering new ways of engineering spatio-temporal photon statistics and potentially realizing tunable sources of correlated quantum light.
\item \textbf{\emph{Driven-dissipative Bose-Hubbard physics.}} In two-dimensional lattices, polaritons offer a route to explore steady-state phases of the driven-dissipative Bose-Hubbard model. Rich phenomena such as multistability, modulational instabilities, and dissipative phase transitions have been predicted~\cite{LeBoite2013, Carusotto2013}. A central challenge is the stabilization of these phases in the presence of losses~\cite{Carusotto2025}, an issue that has recently been addressed in related photonic circuit platforms, where dissipative stabilization of Mott insulating states of light has been experimentally demonstrated~\cite{Ma2019}. Extending such ideas to polariton lattices remains an exciting goal.
\item \textbf{\emph{Strongly correlated topological polaritons.}} Another major direction is the combination of interactions with band topology. Combining polaritonic lattices with strong photon-photon interactions, could lead to the realization of fractional quantum Hall phases, which could be probed through optical spectrocopic measurements~\cite{Umucalilar2012}. Recently, remarkable progress has been made with the observation of Laughlin-like states of photons in different platforms~\cite{Clark2020, Wang2024}. These advances demonstrate the feasibility of preparing and detecting strongly correlated quantum states of light, and motivate ongoing efforts to reach similar regimes in solid-state polariton systems.
\item \textbf{\emph{Bose–Fermi mixtures and superconductivity.}} Finally, several theoretical proposals have investigated the possibility of coupling a polariton condensate to a Fermi sea of electrons, forming a hybrid Bose–Fermi system~\cite{Shelyk2010, Laussy2010, Cotlet2016}. Such systems are predicted to host exotic collective excitations, including roton-like modes, and to mediate effective interactions between electrons, potentially leading to superconductivity with critical temperatures on the order of a few $\unit{\kelvin}$. These proposals highlight the expanded opportunities offered by hybrid light–matter systems to explore diverse quantum many-body phenomena.
\end{itemize}

The quest for polariton quantum matter is still in its early stages, with several technical challenges to address. At the single-site level, enhancing nonlinearities and minimizing losses are essential for achieving strong polariton blockade. In lattices, the goal is to stabilize collective quantum phases under driven-dissipative conditions and, ultimately, to combine strong interactions with lattice engineering. As a complementary approach, circuit QED systems~\cite{Houck2012} provide a versatile platform for exploring strongly interacting photons with precise control over lattice geometry, interactions, and dissipation. Exciton-polaritons, while exhibiting weaker interactions, offer the unique benefit of direct optical access to both the quantum state and its real-time dynamics, enabling measurements and manipulations that may be more challenging in other platforms. Together, these directions point toward a future in which exciton polaritons provide a solid-state platform for exploring quantum many-body physics of light, with potential applications ranging from quantum simulation to quantum metrology and quantum technologies.

\section{Outlook}
\label{Ravets:subsec:5}

In summary, the field of synthetic polariton matter bridges photonics, condensed matter, and quantum many-body physics. Starting from the general motivations of exploring real and synthetic materials, we have seen how lattice engineering in photonic systems allows a precise mapping to tight-binding models, enabling tailored band structures and topological effects. We then discussed how strong photon–photon interactions can be realized \textit{via} exciton-mediated nonlinearities in semiconductor quantum wells, and how the full quantum dynamics of polaritons can be captured using master equation approaches. Finally, we reviewed the rich phenomenology across different regimes: from linear lattice physics and topological effects, to mean-field phenomena such as condensation and superfluidity, and up to the strongly correlated quantum regime. Together, these developments highlight the versatility of the exciton-polariton platform for exploring emergent phenomena in driven-dissipative quantum systems and point toward promising opportunities at the confluence of photonics, topology, and quantum many-body physics.

\begin{acknowledgement}
The author thanks I. Carusotto and M. Wouters for their careful reading of the manuscript and insightful feedback. The author acknowledges funding from the European Research Council (ERC) under the European Union’s Horizon 2020 research and innovation programme through the Starting Grant ARQADIA (grant agreement no. 949730), from the Paris \^Ile-de-France R\'egion via DIM SIRTEQ and DIM QUANTIP, and form the RENATECH network and the General Council of Essonne.
\end{acknowledgement}

\ethics{Competing Interests}{The author has no conflicts of interest to declare that are relevant to the content of this chapter.}

\eject

\end{document}